\documentclass[prb,amssymb,twocolumn,showpacs]{revtex4}

\usepackage{graphicx}
\usepackage{bm}

\begin{document}

\title{Intrinsic leakage of the Josephson flux qubit
and breakdown of the two-level approximation for strong driving}

\author{Alejandro Ferr\'on}
\author{Daniel Dom\'{\i}nguez}

\affiliation{Centro At{\'{o}}mico Bariloche and Instituto Balseiro,
8400 San Carlos de Bariloche,
R\'{\i}o Negro, Argentina.}

\begin{abstract}

Solid state devices for quantum bit computation (qubits) are not 
perfect  isolated two-level systems, since additional higher energy 
levels always exist. One example is the Josephson flux qubit, which 
consists on a mesoscopic SQUID loop with three Josephson junctions 
operated at or near a magnetic flux of half quantum.  
We study intrinsic leakage effects, i.e., direct transitions 
from the allowed qubit states to higher excited states of the system 
during the application of pulses for  quantum computation operations. 
The system is started in the ground state and rf- magnetic field pulses 
are applied at the qubit resonant frequency with pulse   intensity $f_p$.
A perturbative calculation  of the average leakage for small $f_p$ 
is performed for this case, obtaining that the leakage is quadratic 
in $f_p$, and that it depends mainly on the matrix elements of the supercurrent. 
Numerical simulations of 
the time dependent Schr\"odinger equation corresponding to the full 
Hamiltonian of this device  were also performed. 
From the simulations we obtain the value of $f_p$ above 
which the two-level approximation breaks  down, and we estimate
the maximum Rabi frequency that can be achieved.
We study the leakage as a function of the ratio $\alpha$ among the Josephson
energies of the junctions of the device, obtaining the best value 
for minimum leakage ($\alpha\approx0.85$).
The effects of flux noise on the leakage are also discussed.
\end{abstract}


\maketitle

\section{Introduction}

In the last decade different devices of mesoscopic Josephson
junctions  have been studied experimentally as candidates for 
the design of quantum bits (qubits).
\cite{nakamura,qbit_mooij,vion,martinis,revqubits,chiorescu,fqubit_more,noise,mooij09,shimazu} 
A large effort has been devoted to succeed in the 
coherent manipulation of their
quantum states in a controllable way. \cite{nakamura,vion,martinis,chiorescu}
One of the superconductig qubit devices that
has been studied in the last years is the Josephson flux qubit,
\cite{qbit_mooij,chiorescu,fqubit_more,noise,mooij09,shimazu}
which  consists on a mesoscopic SQUID loop with three Josephson junctions 
operated at or near a magnetic flux of half quantum.

Real qubit devices, however, are not perfect isolated two-level systems. 
First, coupling to the
external environment induces relaxation and dephasing.\cite{revqubits,noise} 
Second,  additional higher energy levels always
exist in solid state devices. Therefore leakage effects, i.e.,
transitions from the allowed qubit states to higher excited states
of the system can occur during quantum computation
operations.\cite{fazio99,lloyd00} 
Indirect leakage to the higher energy levels
produced through the interaction with the environment has been
studied in some cases. \cite{burkard,environ-leak} Even neglecting
the interaction with the external environment, 
{\it intrinsic}  leakage
can occur due to direct transitions outside the computational
subspace during the application of pulses for computational
operations.\cite{fazio99,gate-leak,pozzo-lt}
Due to the importance of minimizing the gate errors
due to leakage, several optimization
strategies to compensate leakage, based on varying the
pulse shapes and pulse sequences, 
have been studied recently.\cite{optim-leak-byrd,martinis03,zhou05,rebentrost,martinis08}

Furthermore, the study of the multilevel dynamics of qubit devices has become of
interest by itself in the last years.\cite{buisson,orlando04,lobb,oliver,izmalkov,valenzuela}
The consideration of the superconducting qubit devices as artificial
atoms, has lead to the study of the  dynamic effects of their level
structure beyond the lowest energy levels. 
Effects of strong drive amplitudes on Rabi oscillations have been
studied.\cite{buisson,orlando04,lobb}
Driving the flux qubit with  large amplitude harmonic excitations
have also revealed the higher energy levels through Landau-Zener-Stuckelberg
transitions.\cite{oliver,izmalkov,valenzuela} 
Mach-Zender interferometry\cite{oliver,izmalkov} and
amplitude spectroscopy \cite{valenzuela} have been  the subject of recent studies
of the flux qubit as an artificial atom.
Moreover,  it has also been pointed out that the high energy level structure
of the Josephson flux qubit 
should show  quantum signatures  of classical chaos.\cite{mingo}

For quantum computation applications one wants 
to maximize the number of quantum bit operations before
gate errors become important.
In superconducting qubits,  long pulses are limited by the
decoherence due to the environment
and short pulses by leakage out of the qubit computational subspace.
In order to maximize the number of quantum bit operations,
one has to maximize the
ratio  $t_{\rm deph}/t_{\rm op}$, where $t_{\rm deph}$ is the dephasing time
and $t_{\rm op}$ is the time scale for a single quantum operation. 
The main  approach has been to improve the design
of the qubit devices to increase their coherence time. 
In the case of the Josephson flux qubit
there has been an important progress in increasing  $t_{\rm deph}$, 
from the early experiments by Chiorescu {\it et al.}\cite{chiorescu}
with $t_{\rm deph} \approx 20 {\rm ns}$
to recent experiments that report $t_{\rm deph} \approx 0.5 - 2 \mu{\rm s}$.\cite{noise}
Provided that one has succeeded to achieve a 
$t_{\rm deph}$ as large  as possible for a given device, the following
approach is to reduce $t_{\rm op}$. It is in this later case ($t_{\rm op}\ll t_{\rm deph}$)
when the effect of leakage  is relevant.
The usual strategy for quantum operations is to drive the qubit with
a periodic pulse of intensity $f_p$ at a resonant frequency $\hbar \omega_r=E_1-E_0$,
with $E_1-E_0$ the energy difference between the two qubit states.
In this case, the time scale for quantum computing operations, $t_{op}$,
is proportional to the period of Rabi oscillations,  $t_{op}  \sim T_R$.
Since the time $T_R$ depends on  the pulse strength as $T_R \sim 1 / f_p$,
to reduce $t_{\rm op}$, one has to increase $f_p$.
Along this line of reasoning, 
the following questions will be addressed here: (i) how much is possible to increase 
$f_p$ before leakage effects become important, and 
(ii) for which circuit parameters of the Josephson flux qubit
 the intrinsic leakage is minimum.

To this end, in this work we will 
study the quantum dynamics of the Josephson flux qubit solving
its time dependent Schr\"{o}dinger equation considering the full hamiltonian 
of the system. Since we are interested in time scales 
such that $t_{\rm op}\ll t_{\rm deph}$, the interaction
with the environment will be neglected, and we focus on
the calculation of the amount of intrinsic leakage.
We will study the case when the qubit is driven by
an rf pulse in the magnetic field that is resonant
with $\hbar \omega_r=E_1-E_0$. The amount of leakage
as a function of the pulse strength $f_p$ will
be calculated both perturbatively for small $f_p$ and
numerically for arbitrary values of $f_p$.
The paper is organized as follows.
In Sec.II  we introduce the model hamiltonian
and equations for the Josephson flux qubit.
In Sec. III we present a perturbative calculation
of the leakage in the case of an harmonic resonant
drive.  In Sec.IV we present our numerical results
for the time dependent Schr\"{o}dinger equation, calculating
the amount of leakage as a function of $f_p$. 
In Sec.V we show results on the dependence of te leakage
with different circuit parameters. In particuar, 
the optimum value of the circuit parameter $\alpha$
(the ratio among the Josephson couplings of the junctions
of the qubit) will be computed. In Sec. VI we analyze the
effect of an small amount of noise in the results
shown in the previous section.
Finally, Sec. VII contains a summary 
and a discussion of the most relevant
points of our findings.

\section{Model for the Device for the Josephson Flux Qubit}
\label{model}

\begin{figure}[th]
\begin{center}

\includegraphics[width=0.8\linewidth]{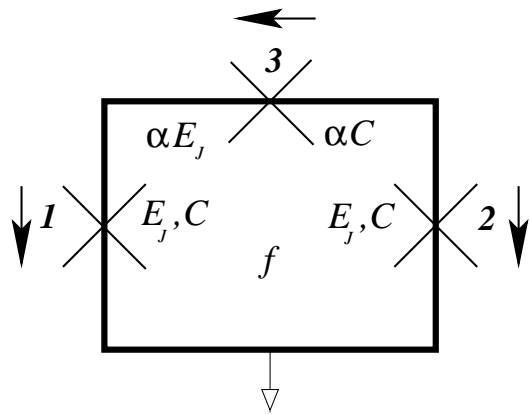}
\caption{Circuit for the Device for the Josephson Flux Qubit as described in the text.
Josephson junctions $1$ and $2$ have Josephson energy $E_J$ and capacitance $C$, and junction $3$
has Josephson energy and capacitance $\alpha$ times smaller.
The arrows indicate the sign
convention for defining the gauge invariant phase differences. The circuit encloses
a magnetic flux $\Phi = f \Phi_0$.} \label{djfq_fig}
\end{center}\end{figure}

The device used for the Josephson flux qubit \cite{qbit_mooij} 
consists on a superconducting ring with  three Josephson junctions   
 enclosing a magnetic flux $\Phi= f\Phi_0$,
with $\Phi_0=h/2e$, see Fig.\ref{djfq_fig}.

The junctions have gauge invariant phase differences defined
as $\varphi_1$, $\varphi_2$ and $\varphi_3$, respectively, with
the sign convention corresponding to the directions
indicated by the arrows in Fig.\ref{djfq_fig}.
Typically the circuit inductance
can be neglected  and the phase difference of the
third junction is:
$\varphi_3=-\varphi _1 +\varphi _2-2\pi f$.
Therefore the system can be described
with  two dynamical variables: $\varphi_1,\varphi_2$.
The circuits that are used for the Josephson flux
qubit have
two of the junctions with the same coupling
energy, $E_{J,1}=E_{J,2}=E_J$, and capacitance, $C_1=C_2=C$,
while the third junction has smaller
coupling $E_{J,3}=\alpha E_J$ and capacitance $C_3=\alpha C$,
with $0.5<\alpha<1$.
The above considerations lead to the Hamiltonian \cite{qbit_mooij}
\begin{equation}
\label{hamil}
{\cal H}=\frac{1}{2}{\vec {P}}^T
{\rm {\bf M}}^{-1}{\vec {P}}
+E_J V(\vec {\bf \varphi})\; ,\label{ham_clas}
\end{equation}
where the two-dimensional coordinate is $\vec{\varphi}=(\varphi_1,\varphi_2)$.
The potential energy term is given by the Josephson energy of the circuit and,
in units of $E_J$, is:
\begin{equation}
\label{eq:pot}
V(\vec {\bf \varphi})=
2+\alpha -\cos \varphi_1-\cos \varphi_2
- \alpha \cos (2\pi f+\varphi _1 -\varphi _2 ) \; .
\end{equation}
The kinetic energy term is given by the electrostatic energy of the circuit, where
the two-dimensional momentum is
$$\vec{P} = (P_1,P_2)={\rm{\bf M}}\cdot \frac{d{\vec{\varphi}}}{dt},$$
and
$\bf M$ is an effective mass tensor determined by the capacitances of the circuit,
$$
{\rm {\bf M}}= C {\left(\frac{\Phi_0}{2\pi}\right)^2} {\rm {\bf m}}\;
$$
with
$$
{\rm {\bf m}}=\left(
{{\begin{array}{cc}
 {1+\alpha  }  & {-\alpha }  \\
 {-\alpha }  & {1+\alpha  }  \\
\end{array} }} \right).$$
We neglected in  $\bf M$ the  on-site capacitances $C_g$ (typically $C_g/C \sim 10^{-2}-10^{-3}\ll 1$).
The system modelled with Eqs.~(\ref{hamil})-(\ref{eq:pot}) is
 analogous to a particle with anysotropic mass ${\rm {\bf M}}$
 in a two-dimensional periodic potential $V(\vec {\bf \varphi})$.

In typical junctions,  the Josephson energy scale, $E_J$, is much larger than
the electrostatic energy of electrons, $E_C= e^2/2C$, and the system
is in a classical regime. On the other hand, mesoscopic junctions (with small area) can have
$E_J\sim E_C$, and quantum fluctuations become important.
In this case, the quantum momentum operator  is defined as
$${\vec {P}} \rightarrow \hat{\vec{P}}= -i\hbar\nabla_\varphi = -i\hbar(\frac{\partial}{\partial\varphi_1},\frac{\partial}{\partial\varphi_2}).$$
After replacing the above defined operator $\hat{\vec{P}}$ in the
Hamiltonian of Eq.(\ref{hamil}), the  eigenvalue Schr\"odinger equation  becomes
\begin{equation}
\label{eq:Schro}
\left[ -\frac{\eta^2}{2}\nabla_\varphi^T{\rm{\bf m}}^{-1}\nabla_\varphi
+V(\vec {\bf \varphi})\right] \Psi_\nu(\vec {\bf \varphi}) = E_\nu \Psi_\nu(\vec {\bf \varphi}) \; ,
\end{equation}
where we normalized energy by $E_J$ and momentum by
$\hbar/\sqrt{8E_C/E_J}$. We defined in Eq.(\ref{eq:Schro})
the parameter $\eta=\sqrt{8E_C/E_J}$ which plays the role of an effective
$\hbar$.\cite{mingo} Typical  flux qubit experiments  have values of $\alpha$ in the range $0.6-0.9$
and $\eta$ in the range $0.1-0.6$.
\cite{chiorescu,fqubit_more,noise,mooij09,shimazu,valenzuela}

In this work, we will study the quantum dynamics of
the Josephson flux qubit. Therefore, we  solve
the time-dependent Schr\"{o}dinger equation which, with the same
normalization as above, is given by
\begin{equation}
i\frac{\partial \Psi(\vec {\bf \varphi})}{\partial t} = \left[
-\frac{\eta^2}{2}\nabla_\varphi^T{\rm{\bf m}}^{-1}\nabla_\varphi
+V(\vec {\bf \varphi})\right] \Psi(\vec {\bf \varphi})\;,
\label{tdse}
\end{equation}
where we normalized time by $t_J
=\hbar/E_J$.

We   integrate
numerically Eq.~(\ref{tdse}) with  a second order split-operator algorithm,
\cite{feit} using a discretization grid of
$\Delta\varphi=2\pi/128$ and $\Delta t= 0.1 t_J$. We use
$2\pi$-periodic boundary conditions on $\vec {\bf
\varphi}=(\varphi_1,\varphi_2)$. Eigenstates   $|\Phi_i\rangle$
and eigenenergies $E_i$ are also calculated by numerical
diagonalization of Eq.~(\ref{eq:Schro}), with the same discretization
grid and boundary conditions.  In what follows
we will consider the case of $\eta=0.48$
({\it i.e.}, $E_J/E_C=35$),  which corresponds to the experiment
of Chiorescu {\it et al.}.\cite{chiorescu}

\section{Perturbative calculation of the intrinsic Leakage}

In quantum computation implementations \cite{qbit_mooij,chiorescu,fqubit_more}
the Josephson flux qubit is operated at magnetic
fields near the half-flux quantum,  $f= 1/2+\delta f$,
with $\delta f \ll 1$.
For values of $\alpha \ge 1/2$, the potential of Eq.(\ref{eq:pot})
has two well defined minima.
At  the optimal operation point $f_0=1/2$, the two lowest energy eigenstates 
($|\Psi_0\rangle$ and $|\Psi_1\rangle$)  are
symmetric and antisymmetric superpositions of two
states corresponding to macroscopic persistent currents
of opposite sign.
A two-level truncation of the Hilbert space is usually performed. \cite{qbit_mooij}
In the subspace expanded by $|\Psi_0\rangle$ and
$|\Psi_1\rangle$, the hamiltonian
of Eq.~(\ref{hamil}) is reduced to
\begin{equation}
 {\cal H}=-\frac{\epsilon}{2} {\hat\sigma}_z - \frac{\Delta}{2} {\hat\sigma}_x
\;,
\end{equation}
where ${\cal H}$ is written in the qubit basis defined by $|0\rangle=(|\Psi_0\rangle+|\Psi_1\rangle)/\sqrt{2}$ 
and $|1\rangle=(|\Psi_0\rangle-|\Psi_1\rangle)/\sqrt{2}$.
Here $\Delta=E_1-E_0$ is
the two-level spliting at $f=1/2$, which increases strongly with $\alpha$,
and  $\epsilon = 4\pi\alpha E_J  S_{01} \delta f$ (considering that $\delta f\ll1$), 
with  $S_{01}=\langle\Psi_0|\sin(\pi +\varphi _1 -\varphi _2)  |\Psi_1\rangle $.
(For typical values of $\alpha$ and $\eta$, one has  $S_{01}\sim 0.8$).
Most experiments control the system varying the magnetic field $\delta f = f-1/2$.
Recently it has been shown experimentally that is also possible to manipulate the value of $\Delta$  
by controlling $\alpha$, replacing the third junction by an additional SQUID loop.\cite{mooij09,shimazu} 

Most of the experiments on the flux qubit study the possibility
of single bit quantum operations by driving the qubit with a  resonant pulse
in the magnetic flux with $\delta f(t)=f_p\sin(\omega_r t)$,
at the resonant frequency $\hbar \omega_r=E_1-E_0$.
If the system is started in the ground state $|\Psi_0\rangle$,
after the pulse is applied during a time interval $\tau$ 
the populations of the ground state and the excited state  are
\begin{eqnarray}
 P_0&=&|\langle\Psi(\tau)|\Psi_0\rangle|^2=
\cos^2(\Omega_R \tau/2)\;, \nonumber \\
P_1&=&|\langle\Psi(\tau)|\Psi_1\rangle|^2=\sin^2(\Omega_R \tau/2)\;,
\nonumber
\end{eqnarray}
with the Rabi frequency $\hbar\Omega_R = \epsilon_p/2$ and 
$\epsilon_p\approx4\pi\alpha  E_J S_{01} f_p $. This result
is usually obtained in the rotating wave approximation (see below). 

In order to check the goodness of the two-level approximation 
we are going to evaluate, perturbatively, the population of the
higher energy levels when a pulse $\delta f(t)=f_p\sin(\omega t)$
is applied, with $f_p\ll 1$. We calculate the leakage outside of the 
quantum computational space spanned by the two lowest levels
as
$$
{\cal L}(t) = \sum_{n=2}^\infty |\langle\Psi(t)|\Psi_n\rangle|^2
= \sum_{n=2}^\infty P_n(t)\;,
$$
where the $|\Psi_n\rangle$ are the eigenstates at $f_0=1/2$.

We now write the hamiltonian of Eq.~(\ref{hamil})
as ${\cal H}= {\cal H}(f_0)+ W(\delta f(t))$ with
$f_0=1/2$ and
\begin{eqnarray}
W(\delta f(t))&=&\alpha E_J\sin[2\pi \delta f(t)]\sin(\pi+\phi_1-\phi_2)+\nonumber\\ 
&&2\alpha E_J\sin^2[\pi \delta f(t)]\cos(\pi+\phi_1-\phi_2)\;,
\end{eqnarray}
for small $f_p$ we have
$$
W(\delta f(t)) \approx 
2\pi f_p\alpha E_J\sin(\omega t)\sin(\pi+\phi_1-\phi_2)\;. 
$$

For the perturbative calculations we  use the fact 
that the first and second states interact strongly with each 
other but only weakly with higher states. In this approximation
we solve (see for example Ref.\onlinecite{merzbacher})

\begin{equation}
i\hbar\frac{\partial c_0(t)}{\partial t}=
W_{00}(t)c_0(t)+W_{01}(t)e^{i\omega_{01}t}c_1(t)\; , \label{pl1}
\end{equation}

\begin{equation}
i\hbar\frac{\partial c_1(t)}{\partial t}=
W_{10}(t)e^{i\omega_{10}t}c_0(t)+W_{11}(t)c_1(t)\;, \label{pl2}
\end{equation}

\begin{equation}
i\hbar\frac{\partial c_n(t)}{\partial t}=
W_{n0}(t)e^{i\omega_{n0}t}c_0(t)+W_{n1}(t)e^{i\omega_{n1}t}c_1(t)\;, \label{pl3}
\end{equation}
with $W_{ij}=\langle \Psi_i|W|\Psi_j\rangle$ and $\omega_{ij}=(E_i-E_j)/\hbar$.

\noindent We rewrite equations (\ref{pl1}) and (\ref{pl2}) in the form 
\begin{equation}
i\frac{\partial c_0(t)}{\partial t}=
\Omega_{00}\sin{(\omega t)}c_0(t)+
\Omega\sin{(\omega t)}e^{-i\omega_{10}t} c_1(t)\,,
\end{equation}

\begin{equation}
i\frac{\partial c_1(t)}{\partial t}=
\Omega\sin{(\omega t)}e^{i\omega_{10}t}c_0(t)+
\Omega_{11}\sin{(\omega t)} c_1(t)\;,
\end{equation}
where we defined $\hbar\Omega_{ij}=2\pi f_p\alpha S_{ij} E_J$ with
$S_{ij}=\langle \Psi_i|\sin(\pi+\phi_1-\phi_2)|\Psi_j\rangle$,
 and $\Omega_{10}=\Omega_{01}=\Omega$ (for $S_{ij}$ real). 
In order to solve these equations we make the 
following change of variables \cite{nik}

\[
b_j(t)=c_j(t)e^{i(\Omega_{jj}/\omega)\cos{(\omega t)}}\, .
\]

\noindent then

\begin{equation}
i\frac{\partial b_0(t)}{\partial t}=
\Omega\sin{(\omega t)}e^{-i(\omega_{10}t+\nu\cos{(\omega t)})}b_1(t)
\label{b0}
\end{equation}

\noindent and

\begin{equation}
i\frac{\partial b_1(t)}{\partial t}=\Omega\sin{(\omega t)}e^{i(\omega_{10}t+\nu\cos{(\omega t)})}b_0(t)\;,
\label{b1}
\end{equation}
\noindent where $\nu=(\Omega_{11}-\Omega_{00})/\omega$. Using the relation 
\[
e^{ix\cos{\theta}}=
\sum_{k=-\infty}^{\infty}i^{k}J_k(x)e^{i n\theta}\;,
\]

\noindent we obtain

\begin{eqnarray}
i\frac{\partial b_0(t)}{\partial t}=\frac{\Omega}{2i}\sum_k i^k J_k(-\nu)\left[e^{-i(\omega_{10}-(k+1)\omega)t}\right.
 \nonumber \\
\left.-e^{-i(\omega_{10}-(k-1)\omega)t}\right] b_1(t)\;,
\label{b02}
\end{eqnarray}

\begin{eqnarray}
i\frac{\partial b_1(t)}{\partial t}=\frac{\Omega}{2i}\sum_k i^k J_k(\nu)\left[e^{i(\omega_{10}+(k+1)\omega)t}
\right. \nonumber\\
\left. -e^{i(\omega_{10}+(k-1)\omega)t}\right] b_1(t)\;,
\label{b12}
\end{eqnarray}

\noindent where the $J_k$ are Bessel functions. 
When $\omega\simeq \omega_{10}$ 
we can use the rotating wave approximation
(RWA) \cite{nik,brown} and negelect the highly off resonant terms obtaining: 

\begin{equation}
\frac{\partial b_0(t)}{\partial t}= -\gamma e^{-i(\omega_{10}-\omega)t}
b_1(t)
\label{b03}
\end{equation}

\noindent and

\begin{equation}
\frac{\partial b_1(t)}{\partial t}= \gamma e^{i(\omega_{10}-\omega)t} b_0(t)\;,
\label{b13}
\end{equation}

\noindent where $\gamma=(\Omega/2)(J_0(\nu)+J_2(\nu))$. 
In the exact resonance case ($\omega=\omega_{10}$) the problem has analytic solution.

\begin{equation}
\frac{\partial b_0(t)}{\partial t}=-\gamma b_1(t) ; \;\;
\frac{\partial b_1(t)}{\partial t}=\gamma b_0(t)\;.
\label{b14}
\end{equation}

\noindent  
If the system was initially in the ground state, we obtain

\begin{equation}
c_0(t)=e^{i(\Omega_{00}/\omega)(1-\cos{(\omega t)})}\cos{(\gamma t)} \label{sol0}
\end{equation}

\noindent and

\begin{equation}
c_1(t)=e^{i(\Omega_{00}/\omega-(\Omega_{11}/\omega)\cos{(\omega t)})}\sin{(\gamma t)}\label{sol1}\, .
\end{equation}

We see that $|c_0(t)|^2=\cos^2(\gamma t)$ and $|c_1(t)|^2=\sin^2(\gamma t)$, which allows
to identify the Rabi frequency as $\Omega_R=2\gamma$, and therefore:
\begin{eqnarray}
\Omega_R&=& \Omega[J_0(\nu)+J_2(\nu)]\nonumber \\
        &=&  \frac{2\pi f_p\alpha S_{01} E_J}{\hbar}\left[J_0(\nu)+J_2(\nu)\right]\;,
                \label{wrabi}
\end{eqnarray}
with $\nu=\frac{2\pi f_p\alpha E_J}{\hbar\omega}(S_{11}-S_{00})$.

\noindent Now we must solve (\ref{pl3}) using the two-level solutions of
Eqs.(\ref{sol0}) and (\ref{sol1}).

\begin{equation}
i\frac{\partial c_n(t)}{\partial t}=
\Omega_{n0}\sin{(\omega t)} e^{i\omega_{n0}t}c_0(t)+
\Omega_{n1}\sin{(\omega t)} e^{i\omega_{n1}t}c_1(t)\;,
 \label{plf}
\end{equation}

\noindent it is easy to show, using (\ref{sol0}) and (\ref{sol1}), that

\begin{eqnarray}
\frac{\partial c_n(t)}{\partial t}&=&
e^{i\Omega_{00}/\omega}\sum_{k=-\infty}^{\infty} i^{k+1}\left[i\beta_{nk0}\left(e^{i a_{nk0}^+t}+
e^{i b_{nk0}^+t}\right. \right. \nonumber \\
&& \left. -e^{i a_{nk0}^-t}-e^{i b_{nk0}^-t}\right)+
\beta_{nk1}\left(e^{i a_{nk1}^+t}+ e^{i b_{nk1}^-t}\right. \nonumber \\
&&-\left.\left.e^{i a_{nk1}^-t}-e^{i b_{nk1}^+t}\right)\right]\;,
\end{eqnarray}

\noindent where $a_{nki}^\pm=k\omega+\omega_{ni}\pm\omega+\gamma$, $b_{nki}^\pm=k\omega+\omega_{ni}\pm\omega-\gamma$
and $\beta_{nki}=(\Omega_{ni}/4)J_{k}(-\Omega_{ii}/\omega)$. 
We  then obtain an expression for the coefficients

\begin{eqnarray}
c_n(t)&=&
e^{i\Omega_{00}/\omega}\sum_{k=-\infty}^{\infty} e^{i(k+1)\pi/2}\left[\beta_{nk0}\left(\frac{e^{i a_{nk0}^+t}}{ a_{nk0}^+}+\frac{e^{i b_{nk0}^+t}}{b_{nk0}^+}\right. \right. \nonumber \\
&& \left. -\frac{e^{i a_{nk0}^-t}}{a_{nk0}^-}-\frac{e^{i b_{nk0}^-t}}{b_{nk0}^-}\right)+
\beta_{nk1}\left(\frac{e^{i a_{nk1}^+t}}{a_{nk1}^+}+ \frac{e^{i b_{nk1}^-t}}{b_{nk1}^-}\right. -\nonumber \\
&&\left.\left.\frac{e^{i a_{nk1}^-t}}{a_{nk1}^-}-\frac{e^{i b_{nk1}^+t}}{b_{nk1}^+}\right)\right]+\delta_{nk0}-i\delta_{nk1}\;,\label{cnt}
\end{eqnarray}

\noindent where $\delta_{nk0}=\beta_{nk0}(1/a_{nk0}^-+1/b_{nk0}^+-1/a_{nk0}^+-1/b_{nk0}^-)$ and
$\delta_{nk1}=\beta_{nk1}(1/a_{nk1}^++1/b_{nk1}^--1/a_{nk1}^--1/b_{nk1}^+)$.

\noindent The average leakage leakage out of the subspace spanned by the first two levels is:

$$
{\bar{\cal L}} = \sum_{n=2}^\infty \overline{|c_n(t)|^2}\;,
$$
where $\overline{|c_n(t)|^2}$ means a time average of $|c_n(t)|^2$.
From (\ref{cnt})  we obtain the perturbative result for the average leakage
as:
\begin{eqnarray}
{\bar{\cal L}} = \sum_{n=2}^\infty
\sum_{k=-\infty}^{\infty} \beta_{nk0}^2\left(\frac{1}{ (a_{nk0}^+)^2}+\frac{1}{(b_{nk0}^+)^2} \right. \nonumber \\
 \left. +\frac{1}{(a_{nk0}^-)^2}+\frac{1}{(b_{nk0}^-)^2}\right)+
\beta_{nk1}^2\left(\frac{1}{(a_{nk1}^+)^2}+ \frac{1}{(b_{nk1}^-)^2}\right. +\nonumber \\
\left.\frac{1}{(a_{nk1}^-)^2}+\frac{1}{(b_{nk1}^+)^2}\right) \nonumber \\
+\left(\sum_{k=-\infty}^{\infty}\delta_{nk0}\cos{(k\pi/2)}+\delta_{nk1}\sin{(k\pi/2)}\right)^2 \nonumber \\ 
+\left(\sum_{k=-\infty}^{\infty}\delta_{nk0}\sin{(k\pi/2)}-\delta_{nk1}\cos{(k\pi/2)}\right)^2\;.\nonumber
\end{eqnarray}

We can simplify the final expression for the leakage taking into account that
$\Omega_{ii}\ll \omega$, since the diagonal matrix elements are $|S_{ii}|\ll 1$.
In this case the term $k=0$ of the Bessel functions is dominant 
in the expansion. We obtain

\begin{equation}
\overline{|c_n(t)|^2}\approx\frac{\Omega_{n0}^2}{\epsilon_{n0}^2}+\frac{\Omega_{n1}^2}{\epsilon_{n1}^2}\;,
\end{equation}

\noindent 
with

\[
\frac{1}{\epsilon_{ni}^2}=\frac{J_0^2(\Omega_{ii}/\omega)}{16}\left(z_{ni}+y_{ni}\right),
\]
and
\[
y_{ni}=\left(\frac{1}{\omega_{ni}-\omega+\gamma}+\frac{1}{\omega_{ni}+\omega-\gamma} \right.
\]
\[
 \left. -\frac{1}{\omega_{ni}+\omega+\gamma}-\frac{1}{\omega_{ni}-\omega-\gamma}\right)^2\;,
\]

\[
z_{ni}=\frac{1}{(\omega_{ni}-\omega+\gamma)^2}+\frac{1}{(\omega_{ni}+\omega-\gamma)^2}+
\]
\[
\frac{1}{(\omega_{ni}+\omega+\gamma)^2}+\frac{1}{(\omega_{ni}-\omega-\gamma)^2}
\;.
\]

\noindent Using that $\hbar\Omega_{ni}= 2\pi f_p\alpha S_{ni}$, with $S_{ni}=\langle\Phi_n|\sin(\pi+\phi_1-\phi_2)|\Phi_0\rangle$,
we finally obtain:

\begin{equation}
{\bar{\cal L}} = \frac{4\pi^2 f_p^2\alpha^2}{\hbar^2} \sum_{n=2}^\infty 
\left(\frac{|S_{n0}|^2}{\epsilon_{n0}^2}+\frac{|S_{n1}|^2}{\epsilon_{n1}^2}\right)\;.
\label{leak_pert}
\end{equation}

\section{Strong driving and breakdown of the two-level approximation}

In this section we solve numerically the time-dependent Schr\"odinger
equation of the full hamiltonian for the Josephson flux qubit, as given
by Eq.~(\ref{tdse}). As a function of time
we calcultate the population of the different energy eigenstates 
when a resonant pulse is applied. In this way we can evaluate directly
how large is the population of the two lowest energy levels and the
amount of leakage as a function of the pulse strength.

We first solve numerically the eigenvalue equation of Eq.~(\ref{eq:Schro})
for $f_0=0.5$, obtaining the eigenvectors $|\Psi_i\rangle$
and eigenvalues $E_i$. Then we solve the corresponding time-dependent Shcr\"odinger
equation of Eq.~(\ref{tdse}). 

We start the time evolution  at the ground state $|\Psi_0\rangle$ 
for $f_0=0.5$. We apply a pulse in the magnetic field, 
$f(t) = f_0 + \delta f(t)$, during a
time interval $\tau$. After the pulse is applied, the wave
function has evolved to $|\Psi(\tau)\rangle$  and the qubit
returns to $f=f_0=0.5$.  We calculate the population $P_i$ of the different eigenstates 
$|\Psi_i\rangle$, obtained at the
end of the pulse: $P_i(\tau)=|\langle \Psi(\tau)|\Phi_i\rangle|^2$. The
leakage outside of the quantum computational subspace expanded by
the two lowest eigenstates, $|\Psi_0\rangle$, $|\Psi_1\rangle$, is
then obtained as 
$${\cal L}(\tau)=1-P_{0}(\tau)-P_{1}(\tau)=\sum_{i=2}^\infty P_{i}(\tau)\;.$$

We consider a resonant rf field pulse, $\delta f(t) =
f_p \sin(\omega_r t)$, for $0<t<\tau$, with the resonant
frequency $\hbar\omega_r= E_1-E_0$. 
In the  experimental measurements of Rabi oscillations of Ref.\onlinecite{chiorescu}, 
pulses of intensity in the range $5\times10^{-5}\lesssim f_p\lesssim
5\times10^{-4}$ were used, as it can be deduced from their data. 
Here we have calculated the time 
evolution of the Schr\"{o}dinger equation in a wider 
range of parameters for the pulse strengths: $10^{-5}<f_p < 0.1$.  


\begin{figure}
\begin{center}
\includegraphics[width=20pc]{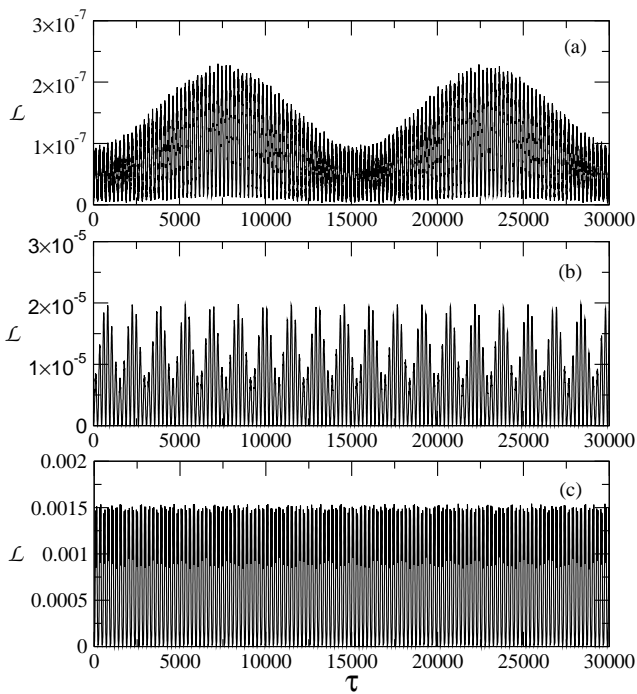}
\end{center}
\caption{Leakage ${\cal L}$ as a function of the pulse length
$\tau$, for an rf field pulse of strength $f_{p}$ at
the resonant frequency $\hbar\omega_r= E_1-E_0$. 
(a) $f_p=0.0001$; (b) $f_p=0.001$; (c) $f_p=0.01$.
Time is normalized by  $t_J =\hbar/E_J\sim 0.5$ps.}
\end{figure}

In this section of the work we are going to present
results for a typical experimental configuration that corresponds
to the work of Chiorescu {\it et al.}\cite{chiorescu}.
In Fig.2 we show ${\cal L}(\tau)$ as a function of the
pulse length $\tau$ for three different cases of $f_p$ for
$\alpha=0.8$. We see that ${\cal L}(\tau)$ has strong oscillations as a function of $\tau$,\cite{comment}
since high frequencies enter into place due to the contribution
of several energy levels.
At $f_p \sim 10^{-4}$, a typical value for experiments,
the average value of the leakage is very small, ${\cal L}\sim
10^{-7}$, showing that under a resonant pulse the Josephson flux qubit
behaves very closely as a two-level system. In constrast, 
a non-resonant pulse can have a higher leakage for similar
pulse strengths, as it has been shown in Ref.\onlinecite{pozzo-lt} 
for a constant dc pulse.
In our results in Fig.2, we see that for increasing values of $f_p$ the 
leakage ${\cal L}(\tau)$ increases, reaching values of $10^{-3}$ for
$f_p \sim 0.01$. A low frequency modulation can be clearly
observed in Fig.2(a). This corresponds to the Rabi frequency of the approximate two-level
system. The two-level Rabi frequency increases with $f_p$. Indeed,  a ``Rabi modulation'' of
the leakage  can also be seen in Fig.2(b) at a higher frequency than in Fig.2(a).
On the other hand, for $f_p=0.01$ in Fig.2(c) the expected
Rabi frequency is high enough that it can not be distinguished from the other
high frequencies contributing to the leakage. As we will see below, for $f_p$ 
above this case the two-level approximation is no longer adequate. 

\begin{figure}
\begin{center}
\includegraphics[width=20pc]{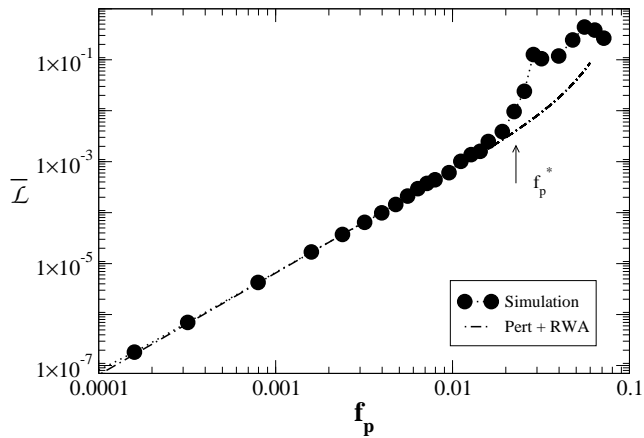}
\end{center}
\caption{Filled circles: Time averaged leakage  ${\bar{\cal L}}$ as a function of
the pulse strength $f_p$ for $\alpha=0.8$ for an rf field pulses of strength $f_{p}$ at
the resonant frequency $\hbar\omega_r= E_1-E_0$. Dash-dotte line: perturbative
approximation of Eq.(\ref{leak_pert}).} 
\end{figure}

To evaluate more quantitatively the effect of strong pulses in
the amount of leakage, we calculate
the time averaged leakage ${\bar{\cal L}}$ as a
function of the pulse intensity $f_p$. We show this result
in Fig.3 for $\alpha=0.8$. We observe that ${\bar{\cal L}}$
grows quadratically with $f_p$ for low pulse strengths.
At large values of $f_p$ the dependence of  ${\bar{\cal L}}$
with $f_p$ clearly departs from this behavior.
We compare in Fig.3 the numerical results with the perturbative
approximation of Eq.(\ref{leak_pert}) (summing up to the first 10 levels). We find that
for $f_p \lesssim f_p^*= 0.02$ the perturbative approximation is
very good. For higher values of the pulse strength, the Eq.(\ref{leak_pert}) 
no longer describes the behavior of ${\bar{\cal L}}(f_p)$, and the the average
leakage increases quickly with $f_p$. 
In particular we find that for
$f_p \sim 0.03$ the amount of leakage is important
(i.e., near $10\%$). From these results, we conclude that 
the two-level approximation can not  be a good description of the dynamics 
for pulse strengths $f_p>f_p^*$.

\begin{figure}
\begin{center}
\includegraphics[width=20pc]{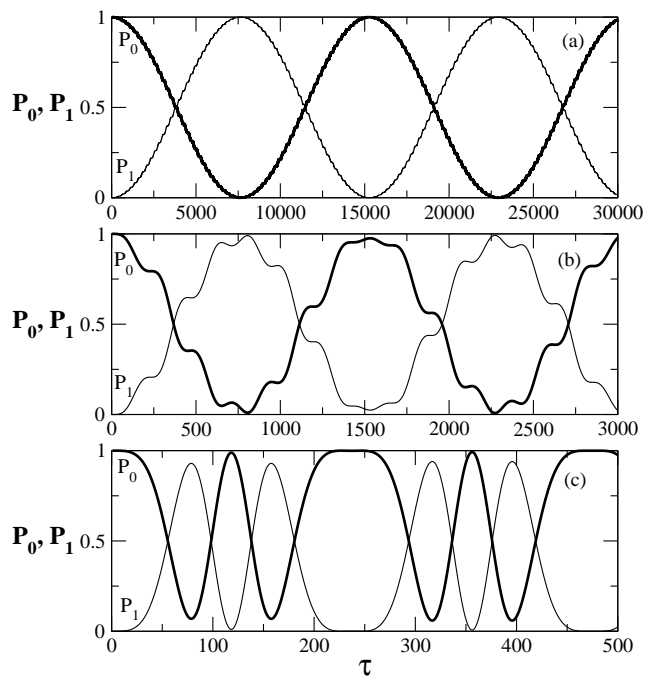}
\end{center}
\caption{Populations of the ground state, $P_0(\tau)$ (thick line), and the
first excited state, $P_1 (\tau)$ (thin line) as a function of the pulse length
$\tau$, for an rf field pulse of strength $f_{p}$ at
the resonant frequency $\hbar\omega_r= E_1-E_0$. (a) $f_p=0.0001$; (b) $f_p=0.001$; (c) $f_p=0.01$.
Time is normalized by  $t_J =\hbar/E_J$.}
\end{figure}

We show in Fig.4 the
population of the ground state $P_0(\tau)$ and the first excited
state $P_1(\tau)$, for the same values of $f_p$ as in Fig.2.
In Fig.4(a), for $f_p=0.0001$, we see a clear Rabi oscillation of the populations
of the two levels of the qubit. A small modulation at high frequencies
is also observed which is due to the perturbation of the higher energy
levels. In Fig.4(b), for $f_p=0.001$, the perturbation of the high energy
levels is more important, but Rabi oscillations can still be 
distinguished. In Fig.4(c), for $f_p=0.01$, we see that the
behavior of $P_0(\tau)$ and  $P_1(\tau)$ departs clearly from the
simple Rabi oscillation scheme and  a more complex oscillation with
two competing frequencies is observed. In this case the
quantum operation of the qubit, if intended, will be more complex
since it needs to be based in the knowledge of the oscillation pattern.

\begin{figure}
\begin{center}
\includegraphics[width=20pc]{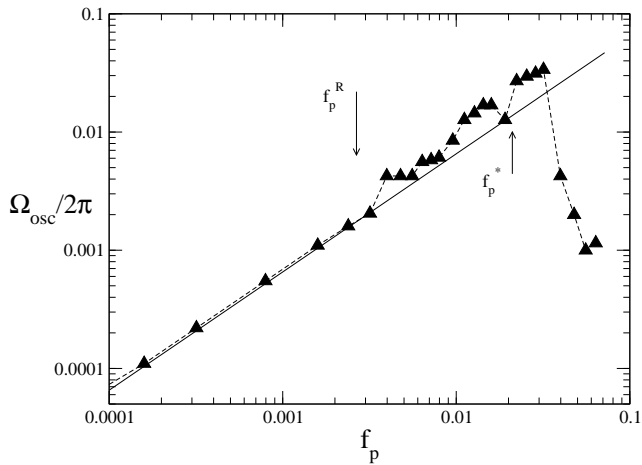}
\end{center}
\caption{Triangles: Main oscillation frequency $\Omega_{\rm osc}$ as a function of
the pulse strength $f_p$ for $\alpha=0.8$. 
Line:  Rabi frequency $\Omega_R$ as obtained in the rotating-wave approximation,
Eq.~(\ref{wrabi})}
\end{figure}

We now analize the Fourier power spectrum of the populations $P_0(\tau)$
and $P_1(\tau)$. For example for $P_0(\tau)$ we calculate
$\hat P(\omega)= 1/T |\int_0^T \exp(i\omega t) P_{0}(t) dt |^2$. 
We define the dominant frequency of
the oscillation from the maximum of the power spectrum,
$\Omega_{\rm osc}={\rm max}_\omega \hat P(\omega)$. A plot 
of the obtained $\Omega_{\rm osc}$ as a function of the pulse
strength $f_p$ is shown in Fig.5. We also plot the Rabi frequency $\Omega_R$
given by Eq.(\ref{wrabi}), which was obtained
in the rotating-wave approximation.
We observe that $\Omega_{\rm osc}$ is
linear with $f_p$ in good agreement with Eq.(\ref{wrabi}), {{it i.e.}  $\Omega_{\rm osc}\approx \Omega_R$  for  $f_p  \lesssim  f_p^R = 0.003$.
For larger values of $f_p$, $\Omega_{\rm osc}$ departs from the
linear dependence of $\Omega_R$. 
The rotating wave approximation that lead to Eq.(\ref{wrabi}) is valid
if $\Omega_R < \omega_r = (E_1-E_0)/\hbar$. Indeed, $\Omega_{\rm osc}$
reaches this value at $f_p^R$ (we see that  for $f_p=f_p^R=0.003$ we have
$\Omega_{\rm osc}/2\pi\approx 0.002$, while $(E_1-E_0)/2\pi = 0.002107$).  
Therefore, for $f_p > f_p^R$ we do not expect to find simple sinusoidal Rabi oscillations.
Instead, more complex oscillations are observed, as seen in Fig.4(c).

From the above analysis we conclude that for $f_p\gtrsim f_p^R = 0.003$ the
possibility of use of the device for quantum operations becomes more difficult
due to the lack of Rabi oscillations.  This implies that the 
highest possible  Rabi frequency that can be obtained
is, for $\alpha=0.8$, 
$$\Omega_R^{max}/2\pi \approx 0.003  \alpha  E_J/\hbar \approx 5 \, \rm{GHz} $$
for $E_J\approx (2\pi\hbar)300$ GHz, and using $\Omega_R \approx 2\pi\alpha f_p E_J$.
In the range $f_p^*>f_p > f_p^R$, the two level approximation is still a good
approximation, since the average leakage is relatively small, but its
use for quantum operations would not be as simple as in the case of $f_p<f_p^R$.
For pulse strengths $f_p  > f_p^*$ the intrinsic leakage is very important (${\cal L}\sim 0.1$)
and the device can not be treated as a two-level system. Moreover, we see in Fig.5
that at $f_p\sim  f_p^*$, the dependence of the frequency $\Omega_{\rm osc}$ with $f_p$
has a drastic change, very far apart from a ``Rabi regime''.

\section{Dependence of the intrinsic leakage with Circuit Parameters}

In this section of the paper we are going to study the behavior of
the  leakage as a function of the different parameters that define
the circuit of the flux qubit.
A simple argument is that leakage effects should be small if the 
energy difference between the third and the second level is much
larger than the energy difference between the two lowest levels,
{\it i.e.} if $E_2-E_1\gg E_1-E_0$. 
While this is generally correct, the magnitude
of the matrix elements for the transition rates to higher energy
levels can be even more important, as we will show in this section.
Indeed, from the perturbative calculation of Eq.(\ref{leak_pert}) we
see that, besides an overall factor proportional to $f_p^2\alpha^2$,  
the average leakage ${\bar{\cal L}}$ depends both on the matrix elements $S_{ni}$,
and the factors $\epsilon_{ni}$ (which are basically dominated by the energy 
level differences $E_n-E_i$).

Among the circuit parameters that can be varied either by external sources
or by circuit design are $f_0$, $\alpha$ and $E_J/E_C$.
As we mentioned before the qubit is operated at a dc magnetic field near the
half-flux quantum,  with $f_0\simeq 1/2$. 
In order to see what happens when we move out of the symmetry point $f_0=1/2$
we plot in Fig.6(a) the time averaged leakage as a function of $f_0$ for a fixed
value of $f_p$. We can see from this figure that the leakage grows as we 
move appart of the symmetry point. This issue can be understood with 
Fig.6(b). Here we can see that if we move out of the symmetry point the 
distance between the two lower levels, $\Delta_{10}=E_1-E_0$ grows  while
the third level becomes closer to the second one, and thus  $\Delta_{21}=E_2-E_1$ decreases.
This basically explains the increase of the leakage when moving out of
the symmetry point $f_0=1/2$. Further information can be obtained
from Fig.6(c)  where we show the matrix 
element $|S_{12}|^2$, which corresponds to the supercurrent 
$\sin(\phi_1-\phi_2-2\pi f)$ taken between the
first excited level and the second excited level. We note that
it has a minimum at $f_0=0.498$. The existence of this
minimum correlates with the fact that the dependence of the leakage with $f_0$
has a shoulder near this point.

\begin{figure}
\begin{center}
\includegraphics[width=20pc]{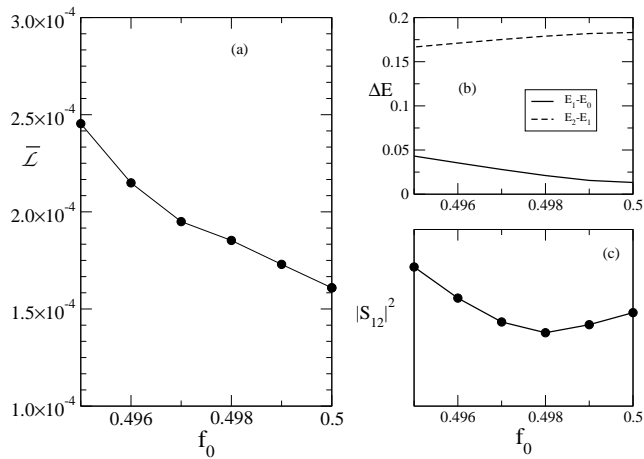}
\end{center}
\caption{(a) Time averaged leakage  ${\bar{\cal L}}$ as a function of
$f_0$ for fixed amplitude $f_p$, $E_J/E_C=35$ and $\alpha=0.8$. (b) Distance
between the lower energy levels as a function of $f_0$. (c) Matrix element
$|S_{12}|^2$ as a function of $f_0$.}
\end{figure}

In principle, the amount of leakage will depend also on the $\alpha$-parameter
since this parameter controls the shape of the effective potential in the
hamiltonian, in particular the heigth of the barrier between the two 
potential minima.  We have calculated numerically the average 
leakage ${\bar{\cal L}}$ as a function of $\alpha$ for a fixed pulse strengths
$f_p$ at $f_0=1/2$. This is shown in Fig.7(a). We see that ${\bar{\cal L}}$ 
tends to decrease with increasing $\alpha$, contrary to what the
$\alpha^2$ factor of Eq.(\ref{leak_pert}) would have suggested.
Moreover, we see that there is a minimum value for the leakage
at $\alpha_{\rm min}\approx 0.85$. Therefore, there is an optimum value of 
the circuit parameter $\alpha$ for which the leakage will be mimimum and 
the two-level approximation more adequate.

\begin{figure}
\begin{center}
\includegraphics[width=20pc]{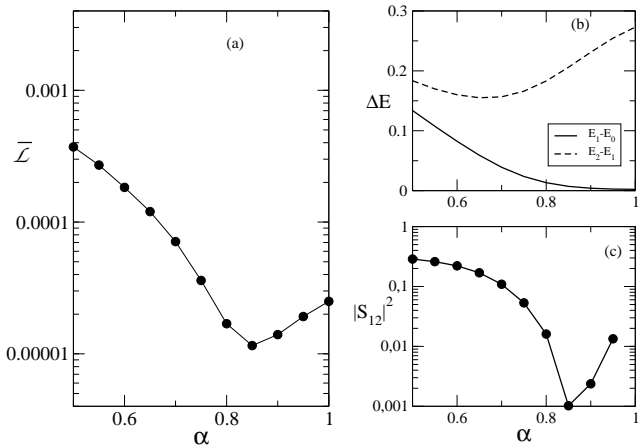}
\end{center}
\caption{(a) Time averaged leakage  ${\bar{\cal L}}$ as a function of
$\alpha$ for fixed amplitude $f_p$, $E_J/E_C=35$ and $f_0=0.5$.
(b) Distance between the lower energy levels as a function of $\alpha$.
(c) Matrix element $|S_{12}|^2$ as a function of $\alpha$.}
\end{figure}

We plot in Fig.7(b) the difference between the three lowest energy levels as a 
function of $\alpha$ and in Fig.7(c) we plot the matrix element $|S_{12}|^2$ vs.  $\alpha$. 
The energy distance $\Delta_{10}$ strongly decreases with $\alpha$, while $\Delta_{21}$
has a smooth non-monotonous dependence with $\alpha$. This later result, where $\Delta_{10}/\Delta_{21}$ decreases with $\alpha$ contributes to the 
general tendency of the leakage to decrease with increasing $\alpha$.
However, the important result here is 
that the minimum leakage at $\alpha_{\rm min}\approx 0.85$ is directly
correlated with a minimum  in the dependence of $|S_{12}|^2$  with $\alpha$, as 
it can be observed in Fig.7(c).
 Therefore, it is this matrix element, for transitions between the second
and the third level, the factor that dominates the dependence of the
leakage with $\alpha$.

\begin{figure}
\begin{center}
\includegraphics[width=20pc]{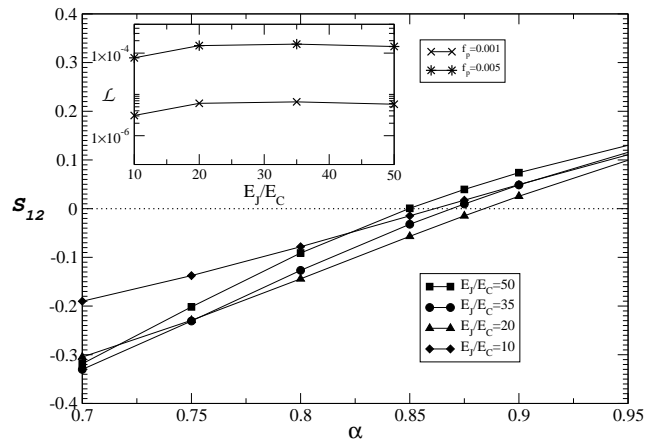}
\end{center}
\caption{Matrix element $S_{12}$ as a function of $\alpha$ for different
values of $E_J/E_C$. The inset shows the time averaged leakage as a 
function of $E_J/E_C$ for two different values of the amplitude.}
\end{figure}

 In Fig.8 we show the amplitude of the matrix element $S_{12}$ as a function of $\alpha$ for
different values of $E_J/E_C$. We observe that the
minimum for $|S_{12}|^2$ shown in Fig.7(c) actually corresponds to the fact
that $S_{12}$ crosses zero and therefore it vanishes at a particular value
of $\alpha$. Furthermore, we find that for all the cases
of $E_J/E_C$ analyzed, the matrix element becomes zero within the range of $0.83<\alpha<0.89$, nearly the same value of $\alpha$. 
In the inset in Fig.8 we show how  the amount of leakage (for a fixed value
of $f_p$, $\alpha$, etc) changes with the ratio $E_J/E_C$.
We see that ${\bar{\cal L}}$ does not change significantly, showing
a slight increase for increasing $E_J/E_C$. This can be
understood in the sense  
the dynamics of the system becomes more ``classical'' when 
increasing $E_J/E_C$, (the effective $\hbar$ decreases), the
energy level structure becomes more crowded,
and thus the effect of higher energy levels will tend to be more relevant.

\section{Effects of  weak noise on the intrinsic leakage}

The results of the previous sections have been obtained in the
ideal case in which the effect of the environment is neglected.
The aim of this section is to analize how  the small perturbation
of a weak noise  can affect the calculations of the intrinsic 
leakage of the previous sections.
In superconducting qubits various sources of relaxation and decoherence are present due
to the environment. Recent experiments in the Josephson flux qubit have
shown that the dominant source of decoherence is due to $1/f$ noise
of the magnetic flux in the SQUID loop.\cite{noise} 
Here we will make the simplification of treating the noise in the magnetic flux
as a classical noise.
In fact, it is usually assumed that for time scales much smaller than the energy relaxation
time, the initial decoherence of the qubit can be described 
with a classical noise.\cite{falci}
Thus, in the presence of an rf-drive, we take the magnetic flux as
\begin{equation}
 f(t) = \frac{1}{2} + f_p \sin(\omega t) + \delta f_n(t) \;, 
\label{fnoise}
\end{equation}
\noindent Here, $\delta f_n(t)$ is the noise source in the magnetic
flux, with average $\langle  \delta f_n(t)\rangle=0$ and correlations $\langle \delta f_n(t) \delta f_n(t') \rangle  = A^2 g(t-t')$.

\begin{figure}
\begin{center}
\includegraphics[width=20pc]{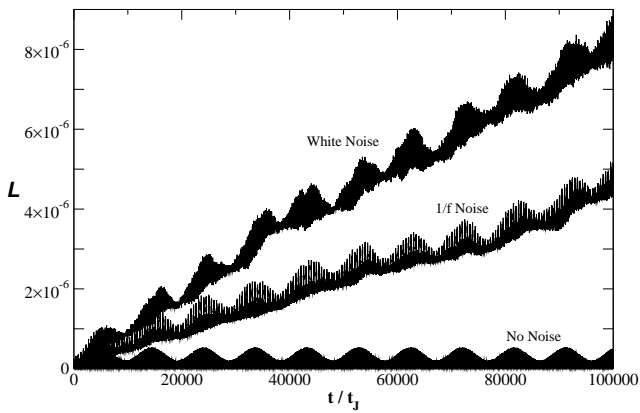}
\end{center}
\caption{Leakage ${\cal L}$ as a function of the pulse length
$\tau$, for an rf field pulse of strength $f_{\rm p}=0.00016$ at
the resonant frequency $\hbar\omega_r= E_1-E_0$ in the presence
of white noise of intensity $A_w=7.10^{-6}$ and $1/f$ noise with $250$ bistable
fluctuators and intensity $A_f=5.10^{-5}$. 
Time is normalized
by  $t_J =\hbar/E_J\sim 0.5$ps.}
\label{n1}
\end{figure}

We now solve numerically the time dependent Schr\"{o}dinger equation, 
Eq.(\ref{tdse}), when the rf-pulse is applied under the presence of noise,
as given by Eq.(\ref{fnoise}).
We calculate the leakage as in the previous section, but now averaging
over $50$ realizations of the noise.
In Fig.\ref{n1} we show the  
leakage ${\cal L}(\tau)$, obtained numerically,  as a function of
the pulse length, for a small value of $f_p$, with Gaussian white noise 
and with  $1/f$ noise. In the first case, we have $g(t-t')=\delta(t-t')$,
and the calculations were done for white noise
intensity $A_w \sim 10^{-5}$. In the second case the $1/f$ noise is
defined as the sum of several bistable fluctuators as studied,
for example, in Ref.~\onlinecite{falci}.
Here we considered $250$ bistable fluctuators with intensity $A_f \sim 10^{-5}$.
As a comparison, in the experiments  the flux noise intensity is
estimated to be of the order of $A\sim 10^{-6}$ at frequencies of $1$Hz.\cite{noise}
In the figure we can see that, besides the oscillations, there is a general linear
increase of the leakage as
$ {\cal L}(t) \approx  \Gamma_L t $, which probably implies an exponential
dependence at long times as  ${\cal L}(t) \approx 1 - \exp(-\Gamma_L t)$. 
As we can see in this figure the case of $1/f$ noise and white noise show
quite similar behavior in their functional form. 
(The intensities of the two noises shown in Fig.\ref{n1}
have been chosen such that they result  in different leak rates $\Gamma_L$ and
thus they can be distinguished in the plot).
The expected difference of the realistic case of $1/f$ noise with a
short-time correlated noise could be in the functional form of the dependence 
of the leakage ${\cal L}(t)$ for large times (larger than the time scale used in the 
present calculation).

We can calculate analytically an approximate result for 
the leakage if we assume that the noise intensity
$A$ is small and that the noise is short-time correlated within a 
small time scale $\tau_n$, such that $A^2t_J\ll \tau_n \ll T_R$, with $T_R$ the
period of the Rabi oscillations.
In Section III we solved the two level system without noise in the 
RWA approximation. Here, we add the effect of noise as a perturbation within 
the RWA approximation. We now write the time dependent perturbation term 
of the Hamiltonian, $W(\delta f(t))$, 
for small values of $f_p$ and small noise intensity, as

\[
W(\delta f(t)) \approx 
2\pi \alpha \sin(\pi+\phi_1-\phi_2) (f_p \sin(\omega t)+\delta f_n(t)). 
\] 

We then solve equation
(\ref{pl3}) with $W_{ni}(t)=2 \pi\alpha S_{ni}E_J(f_p \sin(\omega t)+
\delta f_n(t))$ with the coefficients obtained from the two level approximation
with noise (equations (\ref{pl1}) and (\ref{pl2})).
Integrating equation (\ref{pl3}) and keeping only the terms with $A^2$ 
and $f_p^2$ we obtain

\begin{eqnarray}
 |c_n(t)|^2&=&
|c^{(0)}_n(t)|^2+ \frac{(2\pi\alpha A)^2}{\hbar^2 \gamma}\biggl\{\biggr.\nonumber \\
 && S_{n0} S_{n1} J_0(\Omega_{00}/\omega)J_0(\Omega_{11}/\omega)P(0)\sin^2{(\gamma t)})
 \nonumber \\
&&+\frac{|S_{n0}|^2 J_0^2(\Omega_{00}/\omega)}{4}\Bigl[\gamma(P(\gamma)+P(-\gamma)) t \Bigr.\nonumber\\
&&\Bigl.+P(0)\sin{(2\gamma t)})\Bigr]+ \nonumber \\
&& \frac{|S_{n1}|^2 J_0^2(\Omega_{11}/\omega)}{4}\Bigl[\gamma(P(\gamma)+P(-\gamma)) t \Bigr. \nonumber\\
&&\biggl.\Bigl. -P(0)\sin{(2\gamma t)}\Bigr]\biggr\}\;,
\label{cntn}
\end{eqnarray}

\noindent where $c^{(0)}_n(t)$ is the coefficient obtained in absence of noise in equation 
(\ref{cnt}), and $P(\omega)=\int_{-\infty}^{\infty}e^{i\omega t} g(t) dt$ is the noise
spectral density. We see in Eq.(\ref{cntn}) that, besides the oscillating terms, there are
terms that contribute with
an increase of $|c_n(t)|^2$ which is linear in $t$. This will give an overall increase of the leakage as
${\cal L}(t) = \sum_{n=2}^{\infty} |c_n(t)|^2  \approx  \Gamma_L t \approx 1 - \exp(-\Gamma_L t)$,
which  defines $\Gamma_L$ as the leak rate. 
Typically, in the calculation of relaxation and dephasing, a $1/f$ noise gives
a Gaussian decay law insted of an exponential decay.\cite{noise} Similarly, one could
expect that the exponential dependence of the leakage in the presence of short-time
correlated noise could be modified to a different time dependence in the case of
$1/f$ noise (possibly a quadratic law), but as we showed above, such a difference
is not distinguished in our numerical calculations.

Then, under the assumption of small short-time correlated noise, 
we obtain the leak rate as

\begin{eqnarray}
\Gamma_L&=&\frac{(2\pi\alpha A)^2}{2\hbar^2}\left[P\left(\frac{\Omega_R}{2}
\right)+P\left(\frac{-\Omega_R}{2}\right)\right]\times\label{glp} 
\\ 
&& \left[\sum_{n=2}^{\infty}
|S_{n0}|^2 J_0^2(\Omega_{00}/\omega)
 +|S_{n1}|^2 J_0^2(\Omega_{11}/\omega) \right]\;.\nonumber 
\end{eqnarray}






\begin{figure}
\begin{center}
\includegraphics[width=20pc]{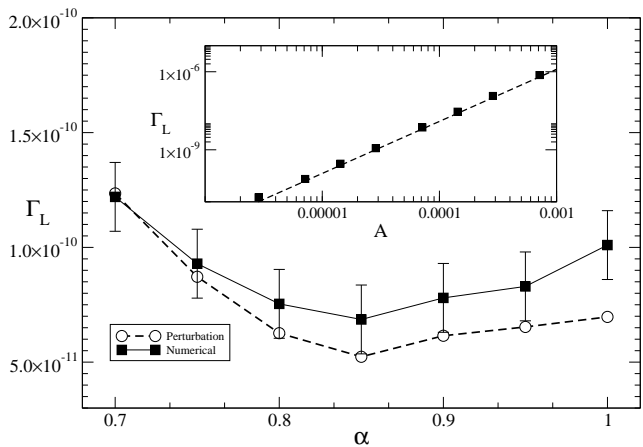}
\end{center}
\caption{Leak rate $\Gamma_L$ as a function of
$\alpha$ for $f_{p}=0.00016$ and white noise strength $A=7\times 10^{-6}$.
Squares: numerical result. Circles: perturbative calculation. Inset: Leak 
rate $\Gamma_L$ as a function of
the white noise strength $A$ for $\alpha=0.8$ and $f_{p}=0.00016$.
Squares: numerical results.
Dashed line: perturbative calculation.}
\label{n2}
\end{figure}

\noindent 
From the numerical results for the white noise, we can make a linear fit of the time 
dependence of ${\cal L}(t)$ to obtain an estimate of the leak rate $\Gamma_L$.
In the inset of Fig.\ref{n2} we show the numerically obtained $\Gamma_L$
as a function of the white noise intensity $A$.  We compare this result 
with the perturbative calculation of Eq.(\ref{glp}). As we can see, 
the agreement is  excelent. 

Finally, for a small value of the white noise intensity $A$, we plot the 
leak rate $\Gamma_L$ as a function of the circuit parameter $\alpha$.
As we can see in Fig. \ref{n2} the leak rate
has a minimum near $\alpha = 0.85$. Therefore
we observe that the optimum value of  $\alpha$ for minimum leakage remains the
same when a small white noise is added in the system. 
We also compare the numerical result with the perturbative
calculation of Eq.(\ref{glp}). It shows an overall agreement with the
existence of a minimum for $\alpha = 0.85$.
We find a small systematic difference for $\alpha > 0.8$.
This can be attributed to the numerical fit procedure used for estimating $\Gamma_L$,
which may need the consideration of larger time intervals to improve the
reliability of the fit.

\section{Summary and Discussion}

We have presented numerical and perturbative calculations of the 
intrinsic leakage in the Josephson flux qubit when the system is driven 
by a rf resonant pulse. 
As a function of the pulse strength $f_p$ three regimes have been found:
(i) For $f_p<f_p^R$, the leakage is very small ${\cal L} \ll 0.001$ and the
device shows good Rabi oscillations. This is the regime in which the device
can be operated as a qubit in a simple way. The perturbative calculations
based on the rotating-wave approximation reproduce very well the numerical
results in this case.
(ii) For $f_p^R<f_p<f_p^*$, the leakage is still small, ${\cal L} < 0.01$,
and the device responds as a two-level system, but the response is more
complex than simple sinusoidal Rabi oscillations. 
(iii) For $f_p>f_p^*$, the leakage is important, ${\cal L} \sim 0.1$ and
the  two-level approximation breaks down.
We also find that the maximum Rabi frequency that can be achieved with
this device, for $f_p=f_p^R$, is about $\Omega_R/2\pi \approx 5 {\rm GHz}$.
This is a factor of $5$-$10$ times larger than what is usually achieved in
typical experiments, meaning that in principle is possible to use stronger
pulses and correspondingly to further reduce the qubit operation time $t_{\rm op}$.

These results  refer to the {\it intrinsic} leakage, {\it i.e.} the leakage
due to direct transitions to the higher energy levels, and therefore 
the effect of the environment has been neglected. The interaction with the environment
also adds further leakage effects due to indirect transitions to 
higher energy levels, see for example Ref.\onlinecite{burkard,environ-leak}. 
Concerning this point, in Sec. VI we have  shown that a weak perturbation
of the external world, considered as a small classical noise, does not change
the qualitative dependence with circuit parameters of the leakage obtained in
the previous sections for the ideal case. These results should be
valid as soon as the Rabi period is much smaller than the decoherence
time, which is the situation analyzed in this paper. 

Here we have considered sharp pulses. Several strategies to reduce leakage
varying the pulse shape and the type of pulse sequences have been discussed 
recently.\cite{optim-leak-byrd,martinis03,zhou05,rebentrost,martinis08}
Applying these strategies the value of the leakage could be further 
reduced from the values obtained here. According to our results, 
the use of these strategies will be of interest in the regime of $f_p < f_p^R$, 
when simple Rabi oscillations can be observed
in the Josephson flux qubit.

One importan result of this work
is the dependence of the leakage with the $\alpha$ parameter
of the device. We have found that there is an optimum value near $\alpha\approx 0.85$
for minimum leakage. This result remains valid when the effect
of noise of small amplitude  in the magnetic flux is considered.
We have found in this case that the magnitude of the leakage is dominated by
the matrix element of the supercurrent for transitions between the first and the second
excited level. When the amplitude of this matrix element crosses zero, 
we find that the leakage is minimum. Moreover, the Fig.7 shows that
small values of $\alpha$ give large leakage, while the range
$0.75\lesssim\alpha<1$ is more favourable. Of course, the 
choice of $\alpha$ in the fabrication of the device should take
into account other factors as dependence with $\alpha$ of the gap, the relaxation
rate, the decoherence time, etc. Interestingly, the experiment of 
Ref.\onlinecite{mooij09} shows that 
the relaxation rate is stronger for small values of $\alpha$, which goes
in the same direction as what we have observed here for the leakage.

Recently, Lucero {\it et al.}\cite{martinis08} have used a procedure (called
``Ramsey filtering'') to measure the population of the second excited level
in the superconducting phase qubit. It will be interesting if a similar
procedure could be implemented in the Josephson flux qubit. In particular,
it will be possible to measure experimentally the dependence of the leakage with $\alpha$
by using ``Ramsey filtering'' in the recently developed circuit
for tuning the gap (i.e. $\alpha$) of the superconducting flux qubit.\cite{mooij09,shimazu}

The results obtained here show quantitatively how the multilevel dynamics
can become relevant for strong driving amplitudes in the flux qubit, going beyond
the two-level approximation.
Indeed, the recent experiments on amplitude spectroscopy of Ref.\onlinecite{valenzuela}
show how the use of strong ac drivings can be turned as a tool to reveal 
the energy level structure of the flux qubit device, now studied
as a solid-sate artificial atom.

\acknowledgments

We acknowledge financial support from ANPCyT (PICT-2007-00824.), CNEA  
and Conicet.


\begin{thebibliography}{}





\bibitem{nakamura} Y. Nakamura, Y. A. Paskin, J. S. Tsai, Nature {\bf 398}, 786 (1999).

\bibitem{qbit_mooij}
J. E. Mooij, T.P.Orlando, L.S. Levitov, L. Tian, C.H. van der Wal and  S. Lloyd,
Science {\bf 285}, 1036 (1999);
T.P.Orlando, J.E. Mooij, L. Tian, C.H. van der Wal, L.S. Levitov, S. Lloyd, J.J. Mazo,
Phys. Rev. B {\bf 60}, 15398 (1999).


\bibitem{vion}  D. Vion, A. Aassime, A. Cottet, P. Joyez, H. Pothier, 
C. Urbina, D. Esteve, and M. H. Devoret,
Science {\bf 296}, 886 (2002).

\bibitem{martinis}
J. M. Martinis, S. Nam, J. Aumentado and C. Urbina, Phys. Rev. Lett. {\bf 89}, 117901 (2002);
 Y. Yu, Y. Yu, S. Han, X. Chu, S.-I. Chu, and Z. Wang, Science {\bf 296}, 889 (2002).

\bibitem{revqubits} Y. Makhlin, G. Sch\"{o}n, and A. Shnirman, Rev. Mod. Phys.
{\bf 73}, 357 (2001).


\bibitem{chiorescu} I. Chiorescu, Y. Nakamura, C. J. P. M. Harmans, and J. E. Mooij,
Science {\bf 299}, 1869 (2003).


\bibitem{fqubit_more} I. Chiorescu, P. Bertet, K. Semba, Y. Nakamura, C. J. P. M. Harmans, and J. E. Mooij,
Nature {\bf 431}, 159 (2004);
E. Il'ichev,  N. Oukhanski, A. Izmalkov, Th. Wagner, M. Grajcar, H.-G. Meyer, A. Yu. Smirnov,
A. Maassen van den Brink, M. H. S. Amin, and A. M. Zagoskin, Phys. Rev. Lett. {\bf 91}, 097906 (2003);
Y. Yu, D. Nakada, J. C. Lee, B. Singh, D. S. Crankshaw, T. P. Orlando, W. D. Oliver, and K. K. Berggren,
Phys. Rev. Lett. {\bf 92}, 117904 (2004);
A. Lupascu, C. J. M. Verwijs, R. N. Schouten, C. J. P. M. Harmans, and J. E. Mooij,
Phys. Rev. Lett. {\bf 93}, 177006 (2004);
P. Bertet, I. Chiorescu, G. Burkard, K. Semba, C. J. P. M. Harmans, D. P. DiVincenzo, and J. E. Mooij,
Phys. Rev. Lett. {\bf 95}, 257002 (2005).



\bibitem{noise} F. Yoshihara {\it et
al.}, Phys. Rev. Lett {\bf 97}, 167001 (2006); K. Kakuyanagi {\it
et al.}, Phys. Rev. Lett. {\bf 98}, 047004 (2007).



\bibitem{mooij09} F. G. Paauw, A. Fedorov, C. J. Harmans, and J. E. Mooij
Phys. Rev. Lett. {\bf 102}, 090501 (2009) 
\bibitem{shimazu} Y. Shimazu, Y. Saito, and Z. Wada
J. Phys. Soc. Japan {\bf 78}, 064708 (2009) 


\bibitem{fazio99} R. Fazio {\it et al.}, Phys. Rev. Lett. 83, 5385
(1999).

\bibitem{lloyd00} L. Tian and S. LLoyd, Phys. Rev.
A, {\bf 62}, 050301 (2000)


\bibitem{burkard} G. Burkard, R. H. Koch, and D. P. DiVincenzo,
Phys. Rev. B {\bf 69}, 064503 (2004).

\bibitem{environ-leak} F. Meier and D. Loss, Phys. Rev. B {\bf 71}, 094519
(2005); T. Hakioglu and K. Savran, Phys. Rev. B {\bf 71}, 115115
(2005).


\bibitem{gate-leak}X. Hu and S. Das Sarma, Phys. Rev. A {\bf 66}, 012312 (2002);
F. Troiani {\it et al.}, Phys. Rev. Lett. {\bf 94}, 207208 (2005).


\bibitem{pozzo-lt} D. Dom\'{\i}nguez and E. N. Pozzo,
 J. Phys.: Conf. Series {\bf 150},   052045 (2009).



\bibitem{optim-leak-byrd} L.-A.Wu, M. S. Byrd, and D. A. Lidar, Phys.
Rev. Lett. {\bf 89}, 127901 (2002); M. S. Byrd {\it et al.}, Phys.
Rev. A {\bf 71}, 052301 (2005).

\bibitem{martinis03} Matthias Steffen, John M. Martinis, and Isaac L. Chuang,
Phys. Rev. B {\bf 68}, 224518 (2003).

\bibitem{zhou05} Z. Zhou {\it et al.}, Phys. Rev. Lett.
{\bf 95}, 120501 (2005).

\bibitem{rebentrost}
P. Rebentrost, and F. K. Wilhelm, Phys. Rev.  B {\bf 79}, 060507 (2009);
F. Motzoi, J. M. Gambetta, P. Rebentrost, F. K. Wilhelm,
Phys. Rev. Lett. {\bf 103}, 110501 (2009).


\bibitem{martinis08} Erik Lucero, M. Hofheinz, M. Ansmann, Radoslaw C. Bialczak, 
N. Katz, Matthew Neeley, A. D. O'Connell, H. Wang, A. N. Cleland, and John M. Martinis,
Phys. Rev. Lett. {\bf 100}, 247001 (2008)





\bibitem{buisson}
J. Claudon, F. Balestro, F. W. Hekking, and O. Buisson 
Phys. Rev. Lett. {\bf 93}, 187003 (2004)
J. Claudon, A. Fay, E. Hoskinson, and O. Buisson
Phys. Rev. B {\bf 76}, 024508 (2007)
J. Claudon, A. Zazunov, F. W. J. Hekking, and O. Buisson 
Phys. Rev. B {\bf 78}, 184503 (2008).


\bibitem{orlando04} 
Yang Yu, D. Nakada, J. C. Lee, B. Singh, D. S. Crankshaw, T. P. Orlando, K. K. Berggren, and W. D. Oliver, Phys. Rev. Lett. {\bf 92}, 117904 (2004)
Z. Dutton, K. V. R. M. Murali, W. D. Oliver, and T. P. Orlando, 
Phys. Rev. B {\bf 73}, 104516 (2006).


\bibitem{lobb}S. K. Dutta, Frederick W. Strauch, R. M. Lewis, Kaushik Mitra, Hanhee Paik, T. A. Palomaki,  Eite Tiesinga, J. R. Anderson, Alex J. Dragt, C. J. Lobb, and F. C. Wellstood
Phys. Rev. B {\bf 78}, 104510 (2008).



\bibitem{oliver} W. D. Oliver, Y. Yu, J. C. Lee, K. K. Berggren,
L. S. Levitov, and T. P. Orlando, Science {\bf 310}, 1653 (2005);
D. M. Berns, W. D. Oliver, S. O. Valenzuela,  A. V. Shytov, K. K. Berggren,
L. S. Levitov, and T. P. Orlando, Phys. Rev. Lett. {\bf 97}, 150502 (2006).

\bibitem{izmalkov} A. Izmalkov, M. Grajcar, E. Il'ichev,  
N. Oukhanski,  Th. Wagner, H.-G. Meyer, W. Krech, M. H. S. Amin,
A. Maassen van den Brink, and A. M. Zagoskin,
Europhys. Lett. {\bf 65}, 844 (2004).


\bibitem{valenzuela} D. M. Berns, M. S. Rudner, S. O. Valenzuela, K. K. Berggren,
W. D. Oliver, L. S. Levitov, and T. P. Orlando, Nature {\bf 455}, 51 (2008);
M. S. Rudner, A. V. Shytov, L. S. Levitov, D. M. Berns, W. D. Oliver, S. O. Valenzuela,and T. P. Orlando,
Phys. Rev. Lett. {\bf 101}, 190502 (2008); 
W D. Oliver  and S. O. Valenzuela, Quantum Inf. Process {\bf 8}, 261 (2009).



\bibitem{mingo} E. N. Pozzo and D. Dom\'{\i}nguez,
Phys. Rev. Lett. {\bf 98}, 057006 (2007); E. N. Pozzo, D.
Dom\'{\i}nguez, and M. J. S\'{a}nchez, Phys. Rev. B {\bf 77},
024518  (2008).










\bibitem{feit} M. D. Feit {\it et al.}, J. Comp. Phys. {\bf 47}, 412 (1982).

\bibitem{merzbacher} E. Merzbacher, {\it Quantum Mechanics}, 2nd edition,
John Wiley \& Sons, 1970, p. 454.

\bibitem{nik}
N. Aravantinos-Zafiris and Emmanuel Paspalakis,
Phys. Rev. A {\bf 72}, 014303 (2005).



\bibitem{brown}
A. Brown, W. J. Meath and P. Tran,
Phys. Rev. A {\bf 63}, 013403 (2000).


\bibitem{comment} In Fig.~2  the intrinsic leakage comes back to the same value
almost periodically in time because  we are consider the dynamics in the ideal
case, neglecting the effect of the environment. The solution of the time dependent
Schrodinger equation always gives almost periodic oscillations of the
populations of the eigenstates. At longer time scales, the effect of the environment
will start to be important, and then the leakage ${\cal L}(\tau)$ should increase with time.
This effect is taken into account later in Fig.~9  in Sec.~VI, in an approximate semiclassical
treatment of noise effects. 

\bibitem{falci} G. Falci, A. D'Arrigo, A. Mastellone, and E. Paladino,
Phys. Rev. Lett. {\bf 94}, 167002 (2005).


 
 
\end{thebibliography}
\end{document}